\documentclass[12pt]{iopart}
\usepackage{graphicx}
\usepackage{color,lscape,fancybox,pifont}
\usepackage{wasysym}

\newcommand{\be}{\begin{equation}}
\newcommand{\ee}{\end{equation}}
\newcommand{\bea}{\begin{eqnarray}}
\newcommand{\eea}{\end{eqnarray}}
\newcommand{\la}{\langle}
\newcommand{\ra}{\rangle}

\begin{document}

\vspace{1.0cm}

\title{  Sharp transition towards shared vocabularies in multi-agent systems} 
\author{Andrea Baronchelli, Maddalena Felici and Vittorio Loreto}
\address{Dipartimento di Fisica, Universit\`a ``La Sapienza'' and SMC-INFM,
P.le A. Moro 2, 00185 Roma, (Italy)}
\author{Emanuele Caglioti}
\address{Dipartimento di Matematica, Universit\`a ``La Sapienza'',
P.le A. Moro 2, 00185 Roma, (Italy)}
\author{Luc Steels}
\address{VUB AI Lab, Brussels (Belgium)}
\address{Sony Computer Science Laboratory, Paris (France)}

\begin{abstract}
What processes can explain how very large populations are able to
converge on the use of a particular word or grammatical construction
without global coordination? Answering this question helps to
understand why new language constructs usually propagate along an
S-shaped curve with a rather sudden transition towards global
agreement. It also helps to analyze and design new technologies that
support or orchestrate self-organizing communication systems, such as
recent social tagging systems for the web. The article introduces and
studies a microscopic model of communicating autonomous agents
performing language games without any central control. We show that
the system undergoes a disorder/order transition, going trough a sharp
symmetry breaking process to reach a shared set of conventions. Before
the transition, the system builds up non-trivial scale-invariant
correlations, for instance in the distribution of competing synonyms,
which display a Zipf-like law.  These correlations make the system
ready for the transition towards shared conventions, which, observed
on the time-scale of collective behaviors, becomes sharper and sharper
with system size. This surprising result not only explains why human
language can scale up to very large populations but also suggests ways
to optimize artificial semiotic dynamics.
\end{abstract}

\maketitle

\section{Introduction}

Bluetooth, blogosphere, ginormous, greenwash, folksonomy.
Lexicographers have to add thousands of new words to dictionaries
every year and revise the usage of many more. Although precise data is
hard to come by, lexicographers agree that there is a period in which
novelty spreads and different words compete, followed by a rather
dramatic transition after which almost everyone uses the same word or
construction~\cite{Lass1997}. This `semiotic dynamics' has lately
become of technological interest because of the sudden popularity of
new web-tools (such as del.icio.us or www.flickr.com) which enable
human web-users to self-organize a system of tags and that way build
up and maintain social networks and share information. Tracking the
emergence of new tags shows similar phenomena of slow spreading
followed by sudden transitions in which one tag overtakes all others.
There is currently also a growing number of experiments where
artificial software agents or robots bootstrap a shared lexicon
without human intervention~\cite{Steels1997, Kirby}. These applications
may revolutionize search in peer-to-peer information
systems~\cite{Steels1998} by orchestrating emergent
semantics~\cite{Staab2002} as opposed to relying on designer-defined
ontologies such as in the semantic web~\cite{Berners-Lee:2001}. They
will be needed when we send groups of robots to deal autonomously with
unforeseeable tasks in largely unknown environments, such as in the
exploration of distant planets or deep seas, hostile environments,
etc.  By definition it will not be possible to define all the needed
communication conventions and ontologies in advance and robots will
have to build up and negotiate their own communication systems,
situated and grounded in their ongoing activities
\cite{Steels2003}. Designers of emergent communication systems want to
know what kind of mechanisms need to be implemented so that the
artificial agents effectively converge towards a shared communication
system and they want to know the scaling laws to see how far the
technology will carry.

\section{The Naming Game}

Some of the earlier work on studying the emergence of communication
conventions has adopted an evolutionary
approach~\cite{Hurford1989,Oliphant1997,NowakKrak1999,Nowak1999,
NowakNiyogiKomarova2001,KomarovaNiyogi2004,NiyogiBerwick1997,Smithetal2003}. Roughly
speaking, the degree in which an agent's vocabulary is similar to that
of others is considered to determine its reproductive fitness, new
generations inherit some features from their parents (vocabularies,
possibly with errors due to their transmission, or learning
strategies), and natural selection drives the population towards
convergence. Here we are interested however in phenomena that happen
on a much more rapid time-scale, during the life-span of agents and
without the need for successive generations. All agents will be
considered peers that have the right to invent and negotiate language
use~\cite{Hutchins95, Steels1995}.  We introduce and study a
microscopic model of communicating agents, inspired by the so-called
Naming Game~\cite{Steels1995}, in which agents have only local
peer-to-peer interactions without central control nor fitness-based
selection, but nevertheless manage to reach a global consensus. There
can be a flux in the population, but generation change is not
necessary for reaching coherence.  Peer-to-peer emergent linguistic
coherence has recently been studied also in~\cite{MatsenNowak2004}
focusing on how a population selects among a set of possible grammars
already known to each agent, whereas here we investigate how
conventions may develop from scratch as a side effect of situated and
grounded communications. The Naming Game model to be studied here uses
as little processing power as possible and thus establishes a
lower-bound on cognitive complexity and performance. In contrast with
other models of language self-organization, agents do not maintain
information about the success rate of individual words and do not use
any intelligent heuristics like choice of best word so far or
cross-situational learning.  We want to understand how the microscopic
dynamics of the agent interactions can nevertheless give rise to
global coherence without external intervention.

\begin{figure}
\vspace{0.5cm}
\begin{center}
\includegraphics[width=7cm]{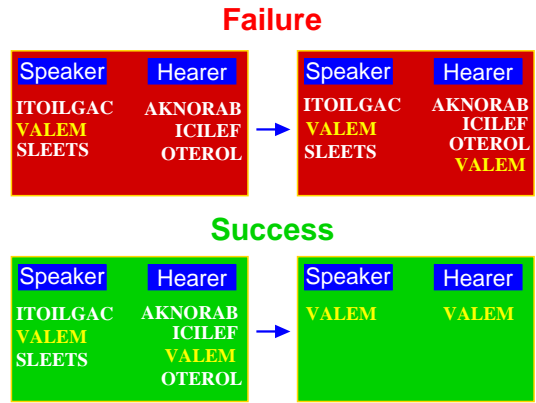}
\caption{{\bf Inventory dynamics:} Examples of the dynamics of the
  inventories in a failed and a successful game, respectively. The speaker
  selects the word highlighted in yellow. If the hearer does not possess that
  word he includes it in his inventory (top). Otherwise both agents
  erase their inventories only keeping the winning word.}
\label{rules}
\end{center}
\end{figure}

The Naming Game is played by a population of $N$ agents trying to
bootstrap a common vocabulary for a certain number $M$ of individual
objects present in their environment, so that one agent can draw the
attention of another one to an object, e.g. to obtain it or converse
further about it.  The objects can be people, physical objects,
relations, web sites, pictures, music files, or any other kind of
entity for which a population aims at reaching a consensus as far
their naming is concerned. Each player is characterized by his
inventory, i.e. the word-object pairs he knows.  All the agents have
empty inventories at time $t=0$. At each time step ($t=1,2,..$) two
players are picked at random and one of them plays as speaker and the
other as hearer. Their interaction obeys the following rules (see
Fig.~\ref{rules}):

\begin{itemize}
\item The speaker selects an object from the current context;
\item The speaker retrieves a word from its inventory associated with the
     chosen object, or, if its inventory is empty, invents a new word;
\item The speaker transmits the selected word to the hearer;
\item If the hearer has the word named by the speaker in its inventory
and that word is associated to the object chosen by the speaker, the
interaction is a success and both players maintain in their
inventories only the winning word, deleting {\bf all} the others;
\item If the hearer does not have the word named by the speaker in its
inventory, the interaction is a failure and the hearer updates its
inventory by adding an association between the new word and the object.
\end{itemize}

\noindent This model makes a number of assumptions. Each player can in
principle play with all the other players, i.e. there is no specific
underlying topology for the structure of the interaction network. So
the game can be viewed as an infinite dimension (or ``mean field'')
Naming Game (an almost realistic situation thanks to the modern
communication networks). Second, we assume that the number of possible
words is so huge that the probability that two players invent the same
word at two different times for two different objects is practically
negligible (this means that homonymy is not taken into account here)
and so the choice dynamics among the possible words associated with a
specific object are completely independent. As a consequence, we can
reduce, without loss of generality, the environment as consisting of
only one single object ($M=1$). In this perspective it is interesting
noting that Komarova and Niyogi~\cite{KomarovaNiyogi2004}, have formally
proven, adopting an evolutionary game theoretic approach, that
languages with homonymy are evolutionary unstable. On the other hand,
it is commonly observed that human languages contain several homonyms,
while true synonyms are extremely rare. In~\cite{KomarovaNiyogi2004}
this apparent paradox is resolved noting that if we think of "words in
a context", homonymy does indeed disappears from human languages,
while synonymy becomes much more relevant. These observations match
perfectly also with our third assumption, according to which speaker
and hearer are able to establish whether the game was successful by
subsequent action performed in a common environment. For example, the
speaker may refer to an object in the environment he wants to obtain
and the hearer then hands the right object. If the game is a failure,
the speaker may point or get the object himself so that it is clear to
the hearer which object was intended.

\section{Phenomenology}

\begin{figure}
\vspace{0.5cm}
\begin{center}
\includegraphics[width=7cm]{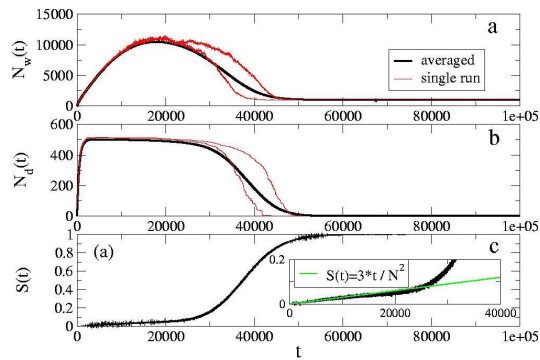}
\end{center}
\caption{{\bf Temporal evolution:} we report here time evolution curves of
  a Naming Game played by $N=1000$ agents. Without loss of
  generality (see text) we consider $M=1$ objects. Bold curves are
  obtained averaging $3000$ runs, while the light ones are obtained by a
  single run. {\bf (a)} Total number of words in the system $N_w(t)$
  vs. $t$ ($t$ here denotes the number of games played) ; {\bf (b)}
  Number of different words in the system $N_d(t)$, whose average
  maximum is $N/2$; {\bf (c)} Success rate $S(t)$, calculated by
  assigning $1$ to a successful interaction and $0$ to a failure and
  averaging over many realizations. In the inset it is shown that, up to
  the disorder/order transition, the success rate is well described by
  the relation $S(t) = 3t/N^2$.}
\label{f_classic}
\end{figure}

The first property of interest is the time evolution of the total
number of words owned by the population $N_w(t)$, of the number of
different words $N_d(t)$, and of the success rate $S(t)$. In
Figure~(\ref{f_classic}) we report these curves averaged over $3000$
runs for a population of $N=1000$ agents, along with two examples of
single run curves. It is evident that single runs originate quite
irregular curves. We assume in these simulations that only two agents
interact at each time step, but the model is perfectly applicable to
the case where any number of agents interact simultaneously.

Clearly, the system undergoes spontaneously a disorder/order
transition to an asymptotic state where global coherence emerges,
i.e. every agent has the same word for the same object. It is
remarkable that this happens starting from completely empty
inventories for each agent. The asymptotic state is one where a word
invented during the time evolution took over with respect to the other
competing words and imposed itself as the leading word. In this sense
the system spontaneously selects one of the many possible coherent
asymptotic states and the transition can thus be seen as a symmetry
breaking transition.

The key question now is whether one can prove that this transition
will always take place and on what time-scale. For our model, it is
easy to prove that an absorbing state will be eventually reached with
probability $1$. Here an absorbing state is a state in which all the
agents have only one word, the same for the whole population. The
proof is straightforward. In fact from any possible state there is
always a non-zero probability to reach an absorbing state in, for
instance, $2(N-1)$ interactions.  A possible sequence is as follows. A
given agent speaks twice with all the other $N-1$ agents using always
the same word (say A). After these $2(N-1)$ interactions all the
agents have only the word A. Denoting with $p$ the probability of the
sequence of $2(N-1)$ steps, the probability that the system has not
reached an absorbing state after $2(N-1)$ iterations is smaller or
equal to $(1-p)$. Therefore, iterating this procedure, the probability
that, starting from any state, the system has not reached an absorbing
state after $2k(N-1)$ iterations, is smaller than $(1-p)^k$ which
vanishes exponentially with $k$. This very general argument, anyway,
does not give any idea about how and on which time scale the absorbing
state is reached. Alternatively one can define the overlap state
function as $\; O=\frac{2}{N(N-1)}\sum_{i>j} \frac{|a_i \cap
a_j|}{|a_i| |a_j|}$, where $a_i$ is the $i^{th}$ agent's inventory,
whose size is $|a_i|$, and $|a_i \cap a_j|$ is the number of words in
common between $a_i$ and $a_j$. The overlap function monitors the
level of lexical coherence in the system.  Averaged over several runs,
it always shows, numerically, a growth with time, i.e. $\la O(t+1) \ra
> \la O(t) \ra $. On the other hand, looking at the single
realization, this function grows almost always, i.e. $\la O(t+1) \ra >
O(t)$ except for a set a very rare configurations whose statistical
weight is negligible. This monotonicity combined with the fact that
the overlap function is bounded, i.e. $O(t) \leq 1$, strongly supports
that the system will indeed reach a final coherent state but a formal
proof is still lacking. This is consistent with the fact that the
coherent state is the only state stable under the dynamical rules of
the model. The more challenging question then concerns under what
scaling conditions convergence is reached.

We can distinguish three phases in the behavior of the system,
compatible with the S-shaped curve typically observed in the spreading
of new language conventions in human
populations~\cite{Lass1997,Best,Korner}. Very early, pairs of agents
play almost uncorrelated games and the number of words hence increases
over time as $N_w(t) = 2t$, while the number of different words
increases as $N_d(t) = t$. In this phase one can look at the system as
a random graph where pairs of agents are connected if they have a word
in common. Because two players always have the same word after a
failed game, each failure at this stage corresponds to adding an edge
to the graph. This fixes a time scale of order $t \sim N$ to establish
a giant component in the network~\cite{percolation1} and for sure
after a time of the order of $t \sim N \log N$ there will be, in the
thermodynamic limit ($N \rightarrow \infty$), only the giant component
surviving~\cite{percolation2}.

Then the system enters a second stage in which it starts building
correlations (i.e. multiple links connecting agents who have more than
one word in common) and collective behavior emerges. We see in the
simulations (see inset of fig.1c) that the rate of success $S(t)$ in
this stage increases as $S(t) \simeq 3t /N^2$ and we have been able to
show analytically why this is the case
\footnote{Details will be reported elsewhere.}.

In this article, we focus on the third stage, when the disorder/order
transition takes place. It occurs close to the time when $N_w(t)$
reaches its maximum. Although one might assume intuitively that the
transition towards global coherence is gradual, we see in fact a
sudden transition towards a consensus, and, even more remarkably, the
transition gets steeper and steeper as the population size
increases. This is important because it shows that the system scales
up to large populations.

\begin{figure}
\vspace{0.5cm}
\begin{center}
\includegraphics[width=7cm]{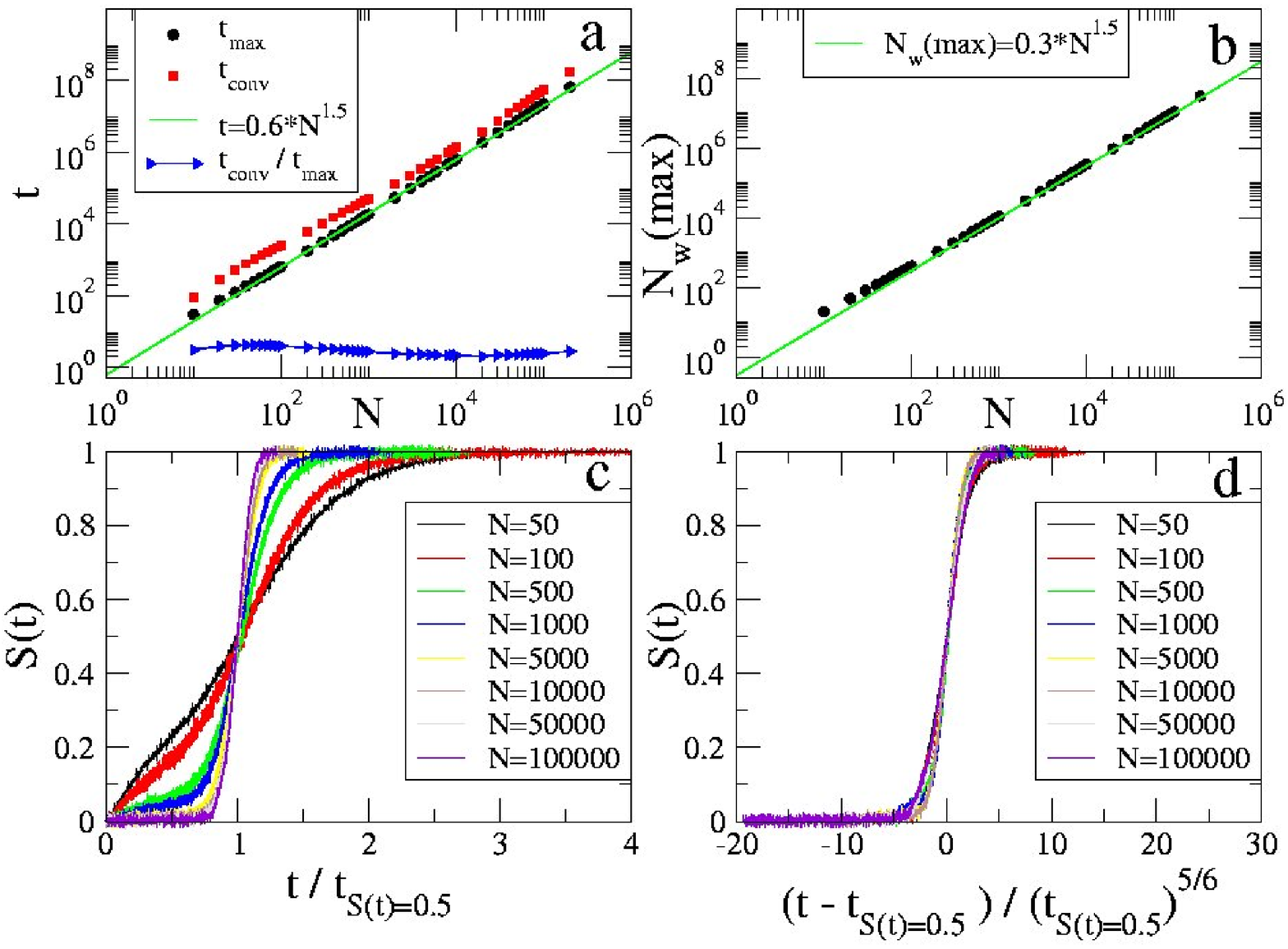}
\end{center}
\caption{{\bf Scaling relations:} {\bf (a)} scaling of the time
  where the total number of words reaches a maximum ($t_{max}$) as well
  as of the convergence times ($t_{conv}$) with the population size
  $N$. Both curves exhibit a power law behavior with exponent
  $3/2$. Statistical error bars are not visible on the scale of the
  graph. An interesting feature emerges from the ratio between
  convergence and maximum times, which exhibits a peculiar oscillating
  trend on the logarithmic scale (mainly due to convergence times
  oscillations). {\bf (b)} scaling of the maximum number of words that
  the system stores during its evolution with the population size
  $N$. The curve exhibits a power law behavior with exponent
  $3/2$. Statistical error bars are not visible on the scale of the
  graph. It must be noted that the values represent the average peak
  height for each size $N$, and this value is larger than the peak of
  the average curve. {\bf (c)} Curves of the success rate $S(t)$ are
  reported for various systems size. The time is rescaled as $t
  \rightarrow (t/t_{S(t)=0.5})$ so that the crossing of all the lines
  at $t/t_{S(t)=0.5}=1$ is an artifact. The increase of the slope with
  system size is evident, showing that the disorder/order transition
  becomes faster and faster for large systems, when the dynamics is
  observed on the system time scale $N^{3/2}$. The form of the
  rescaling has been chosen in order to take into account the
  deviations from the pure power-law behaviour in the scaling of
  $t_{conv}$, rescaling each curve with a self consistent quantity
  ($t_{S(t)=0.5}$).  {\bf (d)} Bottom right: Success rate $S(t)$ for
  various systems size. The curves collapse well after time rescaling
  $t \rightarrow (t-t_{S(t)=0.5})/(t_{S(t)=0.5}^{2/3})^{5/4}$,
  indicating that the characteristic time of the disorder/order
  transition scales as $N^{5/4}$.}
\label{f_scaling}
\end{figure}

\bigskip
\noindent
{\it Time-scales.} In order to better see this phenomenon and then
understand why it is the case, we first look more carefully at the
time scales involved in the process, specifically how the observables
of the system scale with the size $N$ of the
population. Figure~(\ref{f_scaling}a) shows the scaling of the peak
and convergence times of the total number of words with $N$. Both
curves exhibit a power law behavior\footnote{Slight deviation from a
pure power-law behavior are observed for the scaling of the
convergence time. These deviations exhibit a log-periodic behavior and
deserve further investigations.} with an exponent $3/2$. The
distributions for peak and convergence times, for a given size $N$,
are not Gaussian but fit well with the Weibull extreme value
distribution~\cite{gumbel} (data not shown).

The scaling of the maximum number of words $N_w(t_{max})$ is clearly
governed by a power law distribution $N_w(t_{max}) \sim N^{3/2}$ as
well, as shown in Figure~(\ref{f_scaling}b). Here is how the exponent
can be understood using scaling arguments. We assume that, at the
maximum, the average number of words per agent scales as $N^{\alpha}$,
with $\alpha$ unknown. Then it holds:

\be \frac{dN_w(t)}{dt} \propto \frac{1}{c N^{\alpha}} (1 - q) -
\frac{q}{c N^{\alpha}} 2 c N^{\alpha}, \ee

\noindent where, following the model rules, $\frac{1}{c N^{\alpha}}$
is the probability for the speaker to play a specific word. $q$ is he
probability that the hearer possesses the word played by the speaker
which can be estimated as $\frac{c N^{\alpha}}{N/2}$ ($N/2$ being the
number of different words). This is a mean-field assumption since one
neglects the correlations among the inventories and one assumes that
the probability for an agent to possess a given word is
word-independent and is proportional to the number of words in the
agent's inventory. So the two terms are the gain term (in case of a
failed game) and a loss term (in case of a successful game)
respectively where $2 c N^{\alpha}$ (strictly speaking $2 (c
N^{\alpha}-1)$) words are removed from the inventories.

Imposing $\frac{dN_w(t)}{dt}=0$ one gets $\alpha=1/2$. Exploiting the
relation $S(t) \simeq 3t/N^2$ pointed out earlier and valid also at
the the peak, one can predict the scaling of peak time as $t_{max}
\sim N^{\frac{3}{2}}$.

Summarizing, we have a first time scale of the order $N$ where the
system performs uncorrelated language games and the invention process
takes place. It follows the much more interesting time scale
$N^{\frac{3}{2}}$, which is the time-scale for collective behaviors in
the system, i.e. the time scale over which the multi-agent system
collectively builds correlations and performs the spontaneous symmetry
breaking transition.

Figure~(\ref{f_scaling}c) reports success rate curves, $S(t)$, for
different population sizes, all rescaled according to a transformation
equivalent to $t \rightarrow t/N^{3/2}$ (see Figure caption for
details on the rescaling). It is immediately clear that the
qualitative behavior of these curves, when observed on the collective
time-scale $N^{3/2}$, changes with system size $N$. In particular the
transition between a regime of scarce or absent communication, $S(t)
\simeq 0$, and a situation of efficient communication, $S(t) \simeq
1$, i.e. the disorder/order transition, tends to become steeper and
steeper when the population size increases. In order to explain this
phenomenon we need to look at what happens slightly before the
transition.

\section{Network Analysis}

We first investigate the behavior of agent inventories and single
words at the microscopic level. Since each agent is characterized by
its inventory, a first interesting aspect to investigate is the time
evolution of the fraction of players having an inventory of a given
size. A nontrivial phenomena emerges in the fraction of players with 
only one word (data not shown). At 
the beginning, this fraction grows since each player has only one word
after his first interaction, then it decreases, because the first
interactions are usually failures and agents store the new word they
encounter, and eventually it grows again until the moment of
convergence when all the players have the same unique word.  So, the
histogram of the number of agents versus their inventory sizes $k$ is
a precious description of the system at a given time. In particular,
slightly before the convergence, the normalized distribution $p(k)$
deviates from a relatively flat structure to exhibit a power-law
behavior. We can therefore write:

\be
p(k) \sim k^{-{\beta}} f(k/\sqrt{N})
\label{p(k)}
\ee

\noindent with a cut-off function $f(x) = 1$ for $x<< 1$ and
$f(x) = 0$ for $x >> 1$. From simulations it turns out that $\beta
\simeq 7/6$.

\begin{figure}
\vspace{0.5cm}
\begin{center}
\includegraphics[width=7cm]{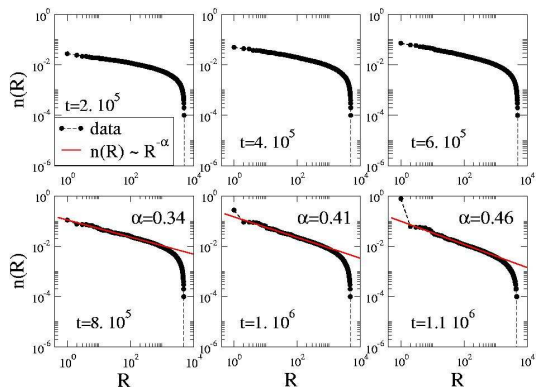}
\end{center}
\caption{{\bf Single words ranking:} The ranking of single words
  is presented for different times for a population size of
  $N=10^4$. The histograms are progressively well described by a power
  law function. For times close to convergence the most popular word
  (i.e. that ranked as 1st) is no more part of the power law trend and
  the whole distribution should be described with eq.~(\ref{n(R)}).}
\label{f_single}
\end{figure}

We now turn to an analysis of the single words themselves. In
Figure~(\ref{f_single}) the different words are ordered according to
their popularity so that the ranking of the most common single word is
$1$. During the first two stages, the distribution of the words can be
described with a power law. However, approaching the transition, the
first ranked single word starts to become more and more popular, while
all the other words are power-law distributed with an exponent
$\alpha$ which changes over time (reminiscent of Zipf's
law~\cite{zipf} and consistent with Polya's urn and other recent
approaches~\cite{polya}). Concretely, the global distribution for the
fraction of agents possessing the $R$-ranked word, $n(R)$, can be
described as:

\be n(R) = n(1) \delta_{R,1} +
\frac{N_w/N-n(1)}{(1-\alpha)((N/2)^{1-\alpha}-2^{1-\alpha})}
R^{-\alpha} f(\frac{R}{N/2}),
\label{n(R)}
\ee

\noindent where the normalization factors have been obtained imposing
that $\int_1^\infty n(R) dR = N_w/N$ \footnote{We substituted the
discrete sums with integrals, an approximation valid in the limit of
large systems.}. On the other hand from equation (\ref{p(k)}) one
gets, by a simple integration, $N_w/N \sim N^{1-\beta/2}$, which gives
$n(R)|_{R > 1} \sim \frac{1}{N^{\beta/2-\alpha}} R^{-\alpha}
f(\frac{R}{N/2})$. This implies that in the thermodynamic limit
$N(1)$, i.e. the number of players with the most popular word, is a
finite fraction of the whole population (a feature reminiscent of the
Bose-Einstein condensation~\cite{Bialas}).

To explain why the disorder/order transition becomes steeper and
steeper in the thermodynamic limit, we must investigate the dynamics
that leads to the state where all agents have the same unique word. In
other words, we need to understand how the network of agents, where
each word is represented by a fully connected clique~\footnote{i.e. a
subset of three or more nodes, with all possible links present.},
reaches its final state of fully connected graph with only single
links. A successful interaction determines the removal of a node from
all the cliques corresponding to the deleted words of the two agents
while a failure causes the increment of an element of the clique
corresponding to the uttered word.  Combining this view of the
population as a network with the fact that the spreading of the most
popular word exceeds that of less common ones, we see that evolution
towards convergence proceeds in a multiplicative fashion, pushing
further the popularity of the most common word while decreasing that
of the others.  An interaction in which the most common word is played
will more likely lead to success, and hence the clique corresponding
to the most common word will tend to increase, while other cliques
will lose nodes. To put this argument on a formal footing, we can
conveniently assume that just before the transition all agents already
know the most popular word. Thus, we have only to determine how the
number of the links deleted after a successful interaction, $M_d$,
scales with $N$, so that we can estimate the rate at which the smaller
cliques disappear from the network. It holds:

\be
M_d = \frac{N_w}{N} \int_{2}^{\infty} n^2(R) N dR \sim N^{3 - \frac{3}{2}\beta}
\label{e_thirdscale}
\ee

\noindent where the product between the average number of words of
each agents (i.e. the average number of cliques involved in each
reduction process), $\frac{N_w}{N}$, the probability of having a word
of rank $R$ (i.e. the probability that the corresponding clique is
involved in the reduction process), $n(R)$, and the number of agents
that have that word (i.e. the size of the clique), $n(R)N$, is
integrated starting from the first deletable word (the second most
popular). From simulations we have that $\beta\simeq 7/6$ so that $M_d
\sim N^{5/4}$ and the ratio $M_d/N^{3/2} \sim N^{-\frac{3}{2}(\beta-1)}
= N^{-1/4}$ goes to zero for large systems. This explains the greater
slope, on the system timescale, of the success rate curves for large
populations (Figure~(\ref{f_scaling}c)). In Figure~(\ref{f_scaling}d)
the time is rescaled as $t \rightarrow (t - \textrm{const.}
N^{3/2})/N^{5/4}$ (see Figure caption for more details on the precise
scaling), and the different $S(t)$ curves collapse indeed well.

\section{Discussion and conclusions}

In this article we have introduced and studied a model of
communication which does not rely on generational transmission
(genetic or cultural) for reaching linguistic coherence but on
self-organization. The model defines the microscopic behavior of the
agents and is therefore directly implementable and thus applicable for
building emergent communication systems in artificial multi-agent
systems. We showed that the model exhibits the same phenomena as
observed in human semiotic dynamics, namely a period of preparation
followed by a rather sharp disorder/order transition. We have
identified the different time-scales involved in the process, both for
individual and collective behaviors. We have explained this dynamics
by observing a build up of non trivial dynamical correlations in the
agents' inventories, which display a Zipf-like distribution for
competing synonyms, until a specific word breaks the symmetry and
imposes itself very rapidly in the whole system.

The Naming Game model studied here is as simple as possible. One can
imagine more intelligent and hence more realistic strategies and the
invention and learning may involve much more complex forms of
language, but that would make the present theoretical analysis less
clear. By focusing on few and simple rules, we have been able to
identify the main ingredients to describe how the population develops
a shared and efficient communication system. The good news, from the
viewpoint of applications, like emergent communication systems in
populations of software agents, is that a well-chosen microscopic
behavior allows a scale-up to very large populations.

{\bf Acknowledgments}~~We thank A. Barrat, L. Dall'Asta, C. Cattuto,
R. Ferrer i Cancho, A. Vulpiani for interesting discussions and a
critical reading of the manuscript. This research has been partly
supported by the ECAgents project funded by the Future and Emerging
Technologies program (IST-FET) of the European Commission under the EU
RD contract IST-1940. The information provided is the sole
responsibility of the authors and does not reflect the Commission's
opinion. The Commission is not responsible for any use that may be
made of data appearing in this publication.

\end{document}